\def\C        {{$^{13}$C \/}}

\newcommand{\mr}[1]{\mathrm{#1}}
\newcommand{\unit}[1]{\,\mathrm{#1}}

\newcommand{\us}{\,\mu{\rm s}}
\newcommand{\uT}{\,\mu{\rm T}}

\newcommand{\rtHz}{\sqrt{\mr{Hz}}}

\newcommand{\To}{T_1}
\newcommand{\Tr}{T_{1\rho}}

\newcommand{\xo}{|0\rangle}
\newcommand{\xp}{|1\rangle}

\newcommand{\xpm}{|\pm1\rangle}

\newcommand{\wo}{\omega_0}
\newcommand{\wone}{\omega_1}

\newcommand{\tc}{\tau_c}
\newcommand{\itc}{\tau_c^{-1}}
\newcommand{\iitc}{\tau_c^{-2}}
\newcommand{\Brms}{B_{\rm rms}}
\newcommand{\SB}{S_B}
\newcommand{\SBB}{S_{\bf B}}
\newcommand{\SBx}{S_{B_x}}
\newcommand{\SBy}{S_{B_y}}
\newcommand{\SBz}{S_{B_z}}

\newcommand{\rhoA}{\rho_A}

\newcommand{\captionstyle}{\normalfont} %define a caption font size

\documentclass[prl,aps,twocolumn]{revtex4}% Phys Rev Lett

\usepackage{graphicx}
\usepackage{bm}% bold math
\usepackage{amsmath}
\usepackage{amssymb}
\usepackage{verbatim}
\usepackage[colorlinks=true, pdfstartview=FitV, linkcolor=blue, citecolor=blue, urlcolor=blue]{hyperref} % enable links

\begin{document}

% formatting
\global\emergencystretch = .1\hsize % adjust the line-breaking to avoid overfull \hbox (standard .15\hsize)

\title{On the surface paramagnetism of diamond}

\author{T. Rosskopf$^{1}$, A. Dussaux$^{1}$, K. Ohashi$^{2}$, M. Loretz$^{1}$, R. Schirhagl$^{1}$, H. Watanabe$^{3}$, S. Shikata$^{4}$, K. M. Itoh$^{2}$, and C. L. Degen$^{1}$}
  \email{degenc@ethz.ch} 
  \affiliation{
   $^1$Department of Physics, ETH Zurich, Schafmattstrasse 16, 8093 Zurich, Switzerland. 
	 $^2$School of Fundamental Science and Technology, Keio University, Yokohama 223-8522, Japan. 
	 $^3$Correlated Electronics Group, Electronics and Photonics Research Institute, National Institute of Advanced Industrial Science and Technology (AIST), Tsukuba Central 4, 1-1-1, Higashi, Tsukuba, Ibaraki 305-8562, Japan. 
	 $^4$Diamond Research Group, Research Institute for Ubiquitous Energy Devices, National Institute of Advanced Industrial Science and Technology (AIST), 1-8-31, Midorigaoka, Ikeda, Osaka 563-8577, Japan. 
	}
\date{\today}

\begin{abstract}
We present measurements of spin relaxation times ($T_1$, $T_{1\rho}$, $T_2$) on very shallow ($\lesssim5$ nm) nitrogen-vacancy (NV) centers in high-purity diamond single crystals.
We find a reduction of spin relaxation times up to 30$\times$ compared to bulk values, indicating the presence of ubiquitous magnetic impurities associated with the surface.
Our measurements yield a density of $0.01-0.1 \mu_B$/nm$^2$ and a characteristic correlation time of $0.28(3)$ ns of surface states, with little variation between samples (implanted, N-doped) and surface terminations (H, F and O).  A low temperature measurement further confirms that fluctuations are themally activated.
The data support the atomistic picture where impurities are associated with the top carbon layers, and not with terminating surface atoms or adsorbate molecules.
The low spin density implies that the presence of a single surface impurity is sufficient to cause spin relaxation of a shallow NV center.
%Our measurements, carried out on two different substrates and for three different surface terminations (hydrogen, oxygen and fluorine), yield a density of $0.01-0.1 \mu_B$/nm$^2$ and a characteristic correlation time of $0.28(3)$ ns of surface spins, with little variation between samples and surface chemistries.  The data support the atomistic picture where impurities are associated with the top carbon layers, and not with terminating surface atoms or adsorbate molecules.  The low spin density implies that the presence of a single surface impurity is sufficient to cause spin relaxation of a shallow NV center.
\end{abstract}

%76.30.Mi 	Color centers and other defects 
%68.35.Dv 	Solid surfaces: Composition, segregation; defects and impurities
%75.70.Cn 	Magnetic properties of interfaces (multilayers, superlattices, heterostructures) 

\maketitle

Interest in magnetic surface impurities of diamond comes from recent attempts to utilize the material for ultrasensitive, nanoscale magnetic sensor heads \cite{degen08, balasubramanian08, maletinsky12} and sensor arrays \cite{toyli10,steinert10,lesage13}.  These sensors take advantage of the long-lived spin state of single nitrogen-vacancy (NV) centers to detect minute magnetic fields down to a few $\unit{nT/\rtHz}$ \cite{maze08,balasubramanian08}.  Diamond-based sensors have recently enabled several notable nanoscale imaging experiments, providing magnetic images of, for example, disk drive media \cite{rondin12,maletinsky12}, magnetic vortices \cite{rondin13}, a single electron spin \cite{grinolds13}, and magnetotactic bacteria \cite{lesage13}.  One of the most exciting prospects of diamond magnetometry is the detection and mapping of single nuclear spins under ambient conditions \cite{degen08}. Such a ``single-spin`` nuclear magnetic resonance (NMR) microscope could have a transformative impact on structural biology and would be an extremely useful tool for the chemical analysis of surfaces.  Indeed, several groups have recently reported successful detection of proton NMR from organic molecules deposited on the surface of a diamond chip with a sensitivity of $10^3-10^6$ nuclei \cite{mamin13,staudacher13,ohashi13}.

Sensitive detection of nuclear spin signals requires placement of NV centers very close to the diamond surface ($<10\unit{nm}$) without compromising the long intrinsic spin coherence time.
Many recent experiments indicate, however, that spin relaxation times of shallow defects can be reduced by several orders of magnitude.  These include studies of nanodiamonds smaller than $50\unit{nm}$ \cite{laraoui12,mcguinness13,tetienne13} and of bulk crystals with NV centers less than about 10 nm from the surface \cite{oforiokai12,mamin12,ohno12,ohashi13}.  The reduction in spin lifetimes is attributed to magnetic noise generated at the diamond surface \cite{panich06,tisler09,oforiokai12}.  A number of possible origins for this noise have been suggested, including dangling bonds \cite{samsonenko79,osipov09}, terminating surface atoms \cite{tisler09,mcguinness13}, adsorbed molecules (like paramagnetic oxygen) \cite{bansal72}, or dynamical strain \cite{flatte13}.  Electron paramagnetic resonance (EPR) and optically detected magnetic resonance (ODMR) have been used to estimate the density $\rhoA$ and characteristic correlation time $\tc$ of surface magnetic states; most studies, however, required extensive modeling, provide ambiguous results, and are not consistent among each other.
Current reported values are between $\rhoA = 0.1-10\unit{\mu_B nm^{-2}}$ and $\tc = 10^{-11} - 10^{-5}\unit{s}$ \cite{mamin12,mcguinness13,tetienne13}.
The goal of this study is to present a clear, quantative picture of the surface density and correlation time, and to more precisely pinpoint the atomistic origin and physical mechanism of noise generation.

%%%%%%%%%%%%%%

Presented are measurements of the spin relaxation times $\To$, $\Tr$ and $T_2$ for a series of shallow ($\lesssim 5\unit{nm}$) NV centers in high-purity, single crystalline diamond.  Relaxation time measurements are widely used techniques in the fields of NMR and EPR for quantitative studies of fast (ps-$\us$) processes in materials \cite{kimmich04,desousa07,schenkel06}.  The methods exploit the fact that the transition rate in a two-level system is proportional to the energy spectral density evaluated at the transition frequency (according to Fermi's golden rule).  For a spin two-level system, the energy spectral density is given by $\gamma^2\SBB(\omega)$, where $\SBB(\omega)$ is the magnetic noise spectral density (in units of T$^2$/Hz) and $\gamma = 2\pi\times28\unit{GHz/T}$ is the electron gyromagnetic ratio.

%Presented are measurements of the spin relaxation times $\To$, $\Tr$ and $T_2$ for a series of shallow ($\lesssim 5\unit{nm}$) NV centers in high-purity, single crystalline diamond. Analysis of relaxation times allows us to directly determine a precise value for the autocorrelation time of surface magnetic fluctuations and to extract an upper bound on the surface spin density. The values found in our study are $\tc=0.28(3)\unit{ns}$ and $\rhoA=0.01-0.1\unit{nm^{-2}}$, with little variation between samples, surface chemistries, and surface environment. These data support the general picture where the surface states are intrinsically associated with diamond's top carbon layers \cite{samsonenko79}, and not with terminating surface atoms or adsorbate molecules.  Our measurements further suggest that the presence of a single surface impurity may be sufficient to cause spin relaxation of a shallow NV center used in magnetometry applications.

%Magnetic resonance relaxometry is a widely used tool in the fields of NMR and EPR for quantitative studies of fast (ps-$\us$) processes in materials \cite{kimmich04,desousa07,schenkel06}.  The method exploits the fact that the transition rate in a two-level system is proportional to the energy spectral density evaluated at the transition frequency (Fermi's golden rule).  For a spin two-level system, the energy spectral density is given by $\gamma^2\SBB(\omega)$, where $\SBB(\omega)$ is the magnetic noise spectral density (in units of T$^2$/Hz) and $\gamma = 2\pi\times28\unit{GHz/T}$ is the electron gyromagnetic ratio.

%
\begin{figure}[t]
\centering
\includegraphics[width=0.41\textwidth]{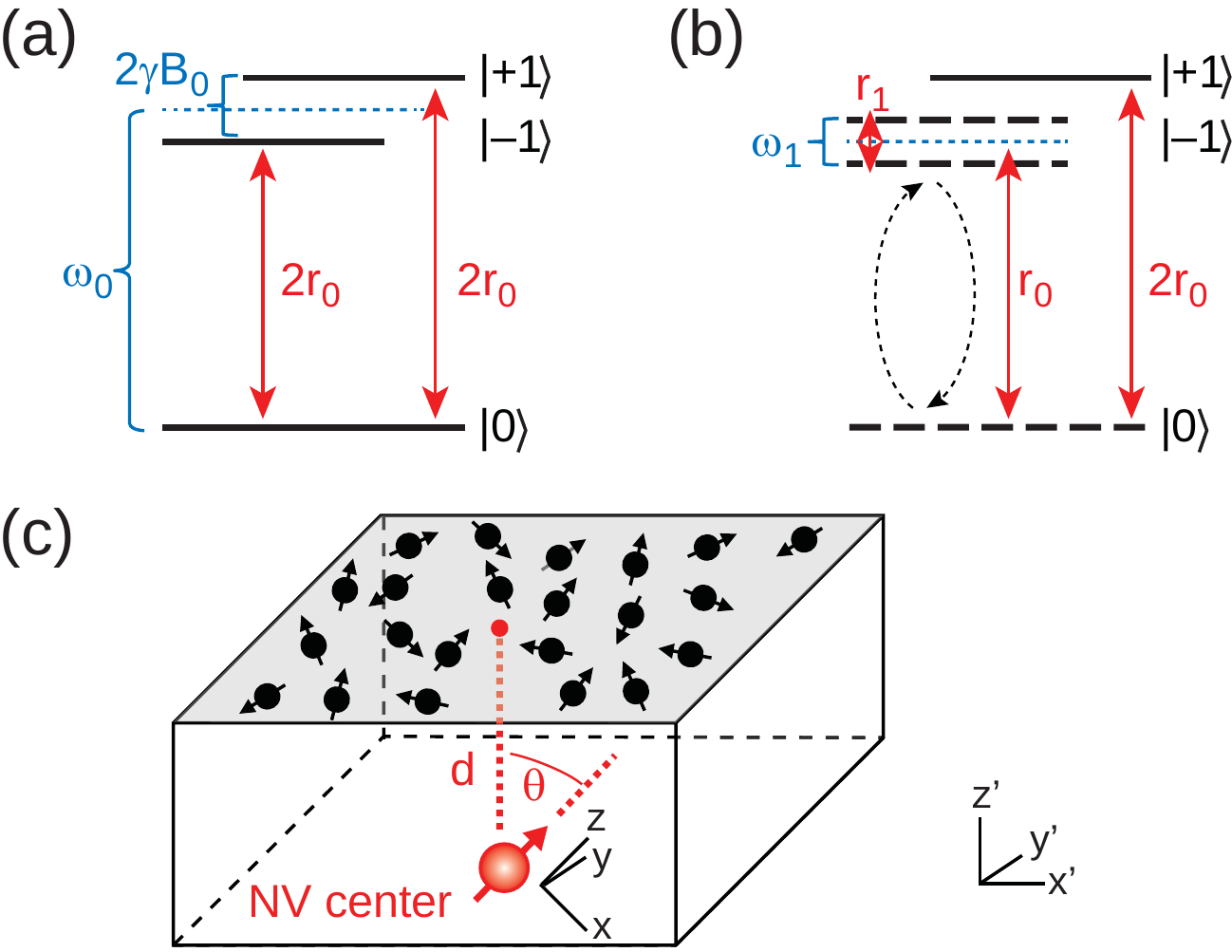}
\caption{\captionstyle
(a,b) Energy level diagram of the nitrogen-vacancy (NV) center's spin $S=1$ system, indicating allowed spin transitions (red solid arrows) and associated transition rates $r_0$ and $r_1$. Blue brackets indicate the energy gaps between states. 
$\wo$ is the Larmor frequency ($\sim2.9\unit{GHz}$) and $\wone$ is the Rabi frequency ($\sim10\unit{MHz}$).  A small bias field ($B_0\sim 10\unit{mT}$) was applied to lift the degeneracy between $|\pm1\rangle$ \cite{jarmola12,biasfield}.  Diagram (a) shows transitions relevant for the $\To$ measurement and diagram (b) shows transitions relevant for the $\Tr$ measurement.  Dashed levels and arrows in (b) symbolize the superposition between $|0\rangle$ and $|-1\rangle$.
(c) Basic picture of diamond surface magnetic impurities. Surface states are represented by a two-dimensional bath of electron spins that produce a fluctuating magnetic field. A nearby shallow NV center is used as a local probe to pick-up the magnetic noise and analyze the spectral characteristics.
}
\label{fig:model}
\end{figure}
Utilized in this study are two different relaxation times that probe $\gamma^2\SBB(\omega)$ on two different time scales, including the spin-lattice relaxation time $\To$ (where $\omega\sim\mathrm{GHz}$) and the rotating-frame relaxation time $\Tr$ ($\omega\mathrm{MHz}$).  Fig. \ref{fig:model}(a) and (b) identify the transitions and rates relevant for these relaxation times in the NV center's spin $S=1$ system.  $r_0$ is the rate of spin flips between the $\xo$ and the (nearly degenerate) $\xpm$ states over an energy gap of $\wo = D = 2\pi\times 2.87\unit{GHz}$, where $D$ is the zero-field-splitting parameter \cite{doherty13}.  $r_1$ is the rate of spin flips between parallel and antiparallel states in a spin-locking experiment \cite{loretz13} with an energy gap given by the Rabi frequency $\wone$ (typ. $\wone\approx 2\pi\times 10\unit{MHz}$).  Transition rates are connected to the magnetic noise spectral density as $r_0 = \frac{1}{12}\gamma^2 \SB(\wo)$ and $r_1 = \frac{1}{12}\gamma^2 \SB(\wone)$ \cite{slichter90,loretz13}, where the numerical factor comes from evaluation of transition matrix elements and $\SB = \SBx + \SBy + \SBz$ is the sum of the three components of the (double-sided) magnetic noise spectral density \cite{orientation}.  The associated relaxation times $\To$ and $\Tr$ are given by (see Supplementary Information):
\begin{eqnarray}
\To^{-1} & = & 6r_0 = \frac{\gamma^2}{2}\SB(\wo) , \label{eq:to} \\
\Tr^{-1} & \approx & 2 r_1 + 3 r_0 = \frac{\gamma^2}{6}\SB(\wone) + \frac{\gamma^2}{4}\SB(\wo), \label{eq:tr} 
\end{eqnarray}
where we have assumed that $\SB(\wone)\gtrsim\SB(\wo)$ and $B_0\ll D$, which will be the case in our study.  Using Eqs. (\ref{eq:to},\ref{eq:tr}) we can express the magnetic noise spectral density as a function of measured $\To$ and $\Tr$:
\begin{eqnarray}
\SB(\wo)   & = & \frac{2}{\gamma^2 \To}, \label{eq:sb0} \\
\SB(\wone) & = & \frac{6}{\gamma^2 \Tr} - \frac{3}{2}\SB(\wo) . \label{eq:sb1} 
\end{eqnarray}

We will interpret the magnetic noise in terms of a two-dimensional bath of electron spins ($S=1/2$) located at a distance $d$ from the NV center, illustrated in Fig. \ref{fig:model}(c).  The two dimensional bath produces a cumulative magnetic field given by the sum of (randomly oriented) magnetic dipoles, and displays a noise spectrum that is governed by the dynamics of the spin bath.  The magnitude of the field is given by:
\begin{equation}
\Brms^2 = \left(\frac{\mu_0}{4\pi}\right)^2
  \iint\limits_{-\infty}^{+\infty} dx' dy' \frac{\rhoA}{r^{6}}
	\sum\limits_{k=x,y,z} \left| \frac{3 {\bf r} ({\bf m}_k\cdot{\bf r})}{r^2} - {\bf m}_k \right|^2 
\end{equation}
where $\rhoA$ is the uniform areal density of surface dipoles, ${\bf r}=(x',y',d)$ is spatial position (with the NV center located at the origin), $r=|{\bf r}|$, ${\bf m}_k$ are the three components of the surface magnetic moment, and $|{\bf m}_k|=\hbar\gamma/2$. For a (100)-oriented surface the NV spin is at $\theta \approx 54.7^\circ$ to the surface normal, and evaluation of the integral yields:
\begin{equation}
\Brms^2 = \frac{3\mu_0^2\hbar^2\gamma_S^2}{64\pi} \times \frac{\rho_A}{d^4} \approx (2.85\,\mr{mT\, nm^3})^2 \times \frac{\rho_A}{d^4} .
\end{equation}
Provided that the depth $d$ of an NV center is known one may use Eq. (\ref{eq:rho}) to infer the density of surface states:
\begin{equation}
\rhoA = \frac{\Brms^2 d^4}{(2.85\unit{mT\, nm^3})^2}. \label{eq:rho}
\end{equation}

The noise spectrum is more difficult to estimate, as it depends on the detailed dynamics of the surface spin bath and may involve multiple time constants.
In spite of that we will interpret dynamics by a single autocorrelation time $\tc$.
%(and note that this in reasonable agreement with earlier susceptibility measurements on different materials surfaces using SQUID's \cite{desousa07,bluhm09}).
The advantage of this approach is that quantitative values for $\tc$ and $\Brms$ can be directly inferred from a single pair of relaxation times, providing an efficient means for analyzing many experimental conditions.  (We note that while the entire noise spectral density could in principle be mapped out by field cycling \cite{kimmich04,loretz13}, these measurements are impractical due to the long acquisition times involved and yield ambiguous results due to the field dependence of surface spin dynamics).  We will find below that intrinsic donor spins in fact contribute additional low frequency ($<$MHz) noise, but this noise is negligible in the high frequency range relevant for this study.  The magnetic noise spectral density associated with correlation time $\tc$ is:
\begin{equation}
\SB(\omega) = \Brms^2 \frac{\itc}{\omega^2+\iitc} , \label{eq:SB}
\end{equation}
which will be used to infer $\tc$ and $\Brms$:
\begin{eqnarray}
\iitc & = & \frac{R\wo^2-\wone^2}{1-R} , \label{eq:itc} \\
\Brms^2 & = & \SB(\wo) \frac{\wo^2+\iitc}{\itc} . \label{eq:brms}
\end{eqnarray}
Here $R = \SB(\wo)/\SB(\wone) = r_0/r_1 \ll 1$ is directly determined by the relaxation times $\To$ and $\Tr$ through Eqs. (\ref{eq:to},\ref{eq:tr}).

%%%%%%%%%%%%%%% Experimental

We have measured spin relaxation times for a series of shallow ($\leq 5\unit{nm}$) NV centers in two different single crystalline samples.  These samples had originally been prepared for other experiments \cite{oforiokai12, ohashi13}, and the data presented here were partially acquired during these measurements.  Sample A was a 17-nm thin film of \C-depleted diamond grown on top of a bulk crystal by chemical vapor deposition (CVD) \cite{ishikawa12,ohashi13,loretz13}.  The topmost 5 nm of this film were doped with nitrogen ($\sim 10$ ppm) during growth, and only this film was found to host NV centers \cite{ohashi13}. Sample B was an electronic-grade single crystal grown by CVD that was scaife-polished, nitrogen-implanted at low energy (0.4-5 keV) and annealed, resulting in NV-centers at roughly 1-10 nm from the surface \cite{oforiokai12}.  Both samples had a (100) surface orientation.  Sample A was further investigated under three different surface chemistries, including hydrogen-, oxygen- and fluorine-terminations.  Sample B was only investigated under oxygen termination.  More details on sample growth and surface preparation are given with Refs. \cite{ohashi13, oforiokai12}.

\begin{figure}[t]
\centering
\includegraphics[width=0.50\textwidth]{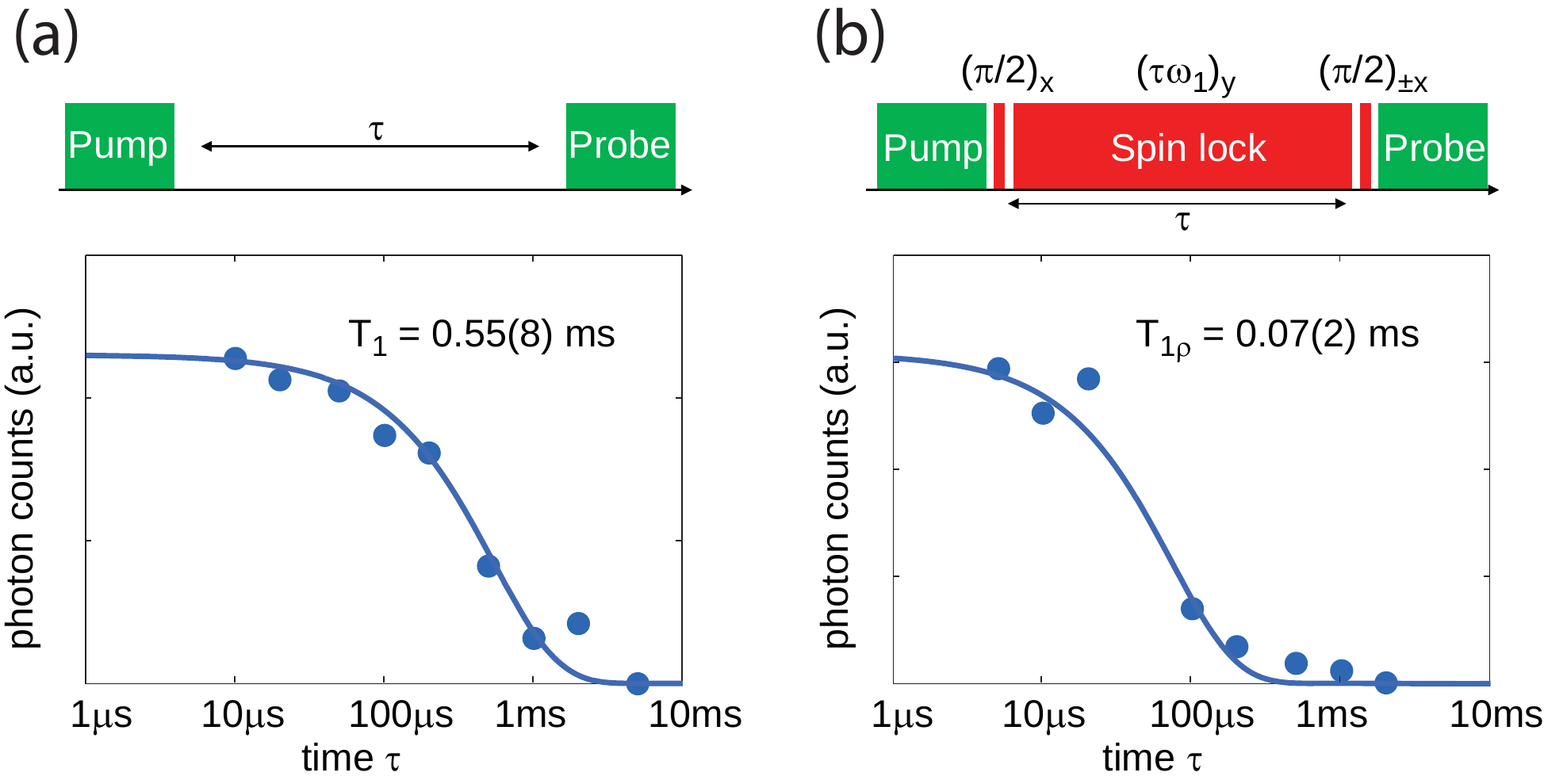}
\caption{\captionstyle  Pulse-timing diagrams and example decay curves for measurements of (a) the spin-lattice relaxation time $\To$ and (b) the rotating frame relaxation time $\Tr$.
Green blocks symbolize laser pulses ($\sim 1\unit{\us}$) and red blocks symbolize microwave pulses.  Blue dots represent experimental values while blue curves are exponential fits.
Data shown is for NV $\#$8 in Fig. \ref{fig:data} measured on sample A.
}
\label{fig:pulsesequence}
\end{figure}
Spin relaxation times were measured by optically-detected magnetic resonance (ODMR) spectroscopy at room temperature \cite{gruber97}.  All experiments were carried out on single NV centers.  Measurement protocols are explained in Fig. \ref{fig:pulsesequence}:  For $\To$ measurements, NV centers were prepared by a first ``pump'' laser pulse into the $\xo$ state and the spin state was measured by a second ``probe'' laser pulse based on the NV center's spin-dependent luminescence \cite{gruber97}.  Pump and probe pulses were separated by a dark interval of duration $\tau$ during which relaxation occurred.  Mapping of fluorescence generated by the probe pulse as a function of $\tau$ then yields an exponential decay with decay time $\To$.  Two decay curves were recorded for each NV center with the spin state initialized in the $\xo$ and $\xp$ state, respectively, to obtain, after subtraction, a zero-baseline measurement with a monoexponential decay \cite{jarmola12}.
For $\Tr$ measurements, three microwave pulses were applied during the dark period to create a ``spin lock'' situation \cite{loretz13} (see Fig. \ref{fig:pulsesequence}(b)).  Again, phase cycling of the second pulse was used to achieve a zero-baseline measurement with a monoexponential decay.  Representative decay curves for $\To$ and $\Tr$ are shown with Fig. \ref{fig:pulsesequence}.

%%%%%%%%%%%%%%% Results

\begin{figure}[t]
\centering
\includegraphics[width=0.42\textwidth]{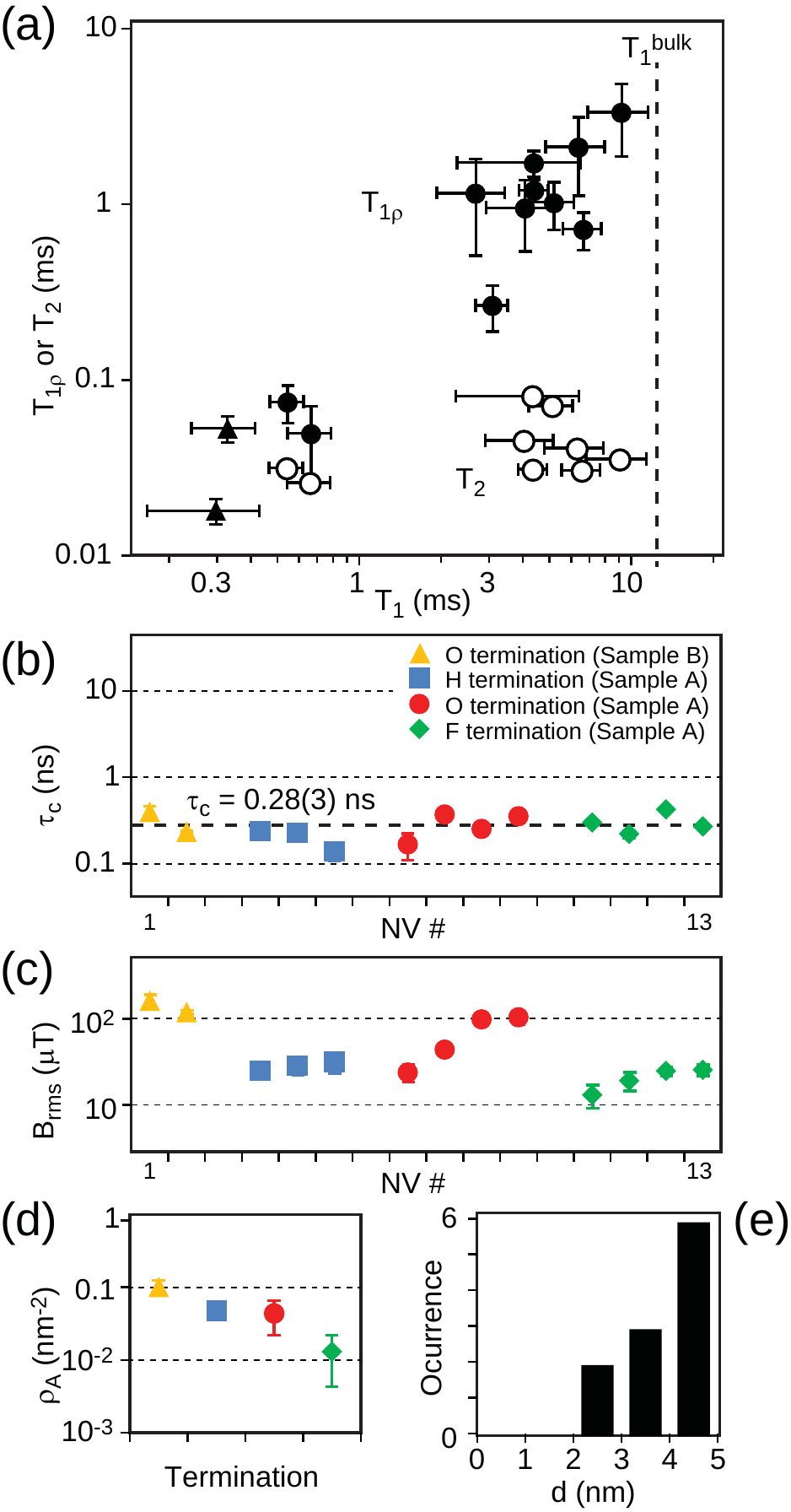}
\caption{\captionstyle
Spin relaxation time measurements of 13 shallow NV centers for different samples and surface terminations.
(a) $\To$, $\Tr$ and $T_2$ times organized in a two-dimensional correlation plot. Observed relaxation times are $\To=0.3-9\unit{ms}$, $\Tr=0.02-3\unit{ms}$ and $T_2=22-80\unit{\us}$. Dots represent Sample A and triangles represent Sample B. Bulk $\To^\mr{bulk} = 12(2)\unit{ms}$ (Sample A) is indicated by a dashed line.
(b) Autocorrelation time $\tc$ of surface fluctuations for the same NV centers, organized by sample and surface termination.
(c) rms magnetic field $\Brms$ for each NV center.
(d) Density of surface impurities $\rhoA$ for samples and surface terminations.
(e) Histogram of depth values (upper bound) of NV centers inferred from $\Brms$.
Errors are propagated from fits to $\To$ and $\Tr$ decay curves.  Numerical data are provided as Supplementary Information.
}
\label{fig:data}
\end{figure}

Fig. \ref{fig:data} collects and analyzes measurements obtained from 13 different NV centers.  These measurements represent the main dataset of our study.  In a first panel (Fig. \ref{fig:data}(a)) we plot $\Tr$ as function of $\To$ (black dots).  The figure serves to illustrate two findings: To begin, we observe that $\Tr$ and $\To$ are strongly correlated -- NV centers with long $\To$ times also have long $\Tr$, and NV centers with short $\To$ times also exhibit short $\Tr$.  The ratio between $\To$ and $\Tr$ is fairly consistent at about 10:1.  Second, we observe that relaxation times are reduced up to $30\times$ compared to bulk values (here $\To^\mr{bulk} = 12(2)\unit{ms}$ for Sample A and $\To^\mr{bulk} = 4.5(2)\unit{ms}$ for Sample B).  This shows that surface effects are indeed present and that both $\To$ and $\Tr$ are sensitive indicators of surface magnetic noise.  The baseline noise associated with $\To^\mr{bulk} = 12\unit{ms}$ of sample A is only $\SB = 2/(\gamma^2 \To^\mr{bulk}) = 73\unit{pT/Hz^{1/2}}$, illustrating the sensitivity of the measurement. 

Fig. \ref{fig:data}(a) additionally plots values of the spin echo decay time $T_2$.  We note that $T_2$ is not correlated with $\To$.  Thus, $T_2$ relaxation is governed by low-frequency ($\sim\mathrm{kHz}$) noise that is not related to the surface, such as the noise produced by nitrogen impurities in the diamond samples.  $T_2$ values measured in this study were between $10-100\unit{\us}$.

Figs. \ref{fig:data}(b) and (c) plot values of the characteristic correlation time $\tc$ and the rms magnetic field $\Brms$ organized by surface chemistry and sample type.  $\tc$ and $\Brms$ were calculated from relaxation times according to Eqs. (\ref{eq:sb0},\ref{eq:sb1}) and (\ref{eq:itc},\ref{eq:brms}). (We note that an offset $\SB = 2/(\gamma^2 \To^\mr{bulk})$ was subtracted from both $\SB(\wo)$ and $\SB(\wone)$ to account for surface-unrelated or ``bulk'' relaxation).  We find that $\tc$ shows little variation with most values between 0.2 and 0.4 ns, around a mean value of $\tc = 0.28(3)\unit{ns}$.  This finding is surprising, because a strong variation of $\tc$ would be expected if magnetic surface states were rooted in terminating surface atoms or adsorbates.  Much larger variations are found for $\Brms$, as can be expected from the stochastic placement of NV centers and perhaps also due to the stochastic distribution of surface impurities.

In Fig. \ref{fig:data}(d) we have calculated an upper bound for the surface spin density $\rhoA$ based on $\Brms$ and an estimate of defect depth $d$.  Although we do not have a precise knowledge of $d$, we know that $d\lesssim 5\unit{nm}$ for all NV centers given the 5-nm-thick doping layer of the sample (Sample A) \cite{depth}. For sample B the depth was estimated through SRIM calculations \cite{oforiokai12,srim}.  Among the NV centers of Sample A we have pick the ones with the lowest $\Brms$ for each surface termination (here $\sim 20\unit{\uT}$). These NV centers are likely bvery close to $d=5$ nm.  In fact, since many NV centers showed similar $\Brms\sim 20\unit{\uT}$, we suspect that most NV centers are located near the deep end of the doping layer.  We find that $\rhoA = 0.01-0.1\unit{nm^{-2}}$ (upper bound) for both samples (see Fig. \ref{fig:data}(d)). The lowest densities are observed for fluorine-terminated surfaces and the highest densities for the implanted surface, respectively.

The densities of surface impurities found here are low compared to previous studies on nanodiamonds \cite{tisler09,tetienne13} and compared to densities measured by SQuID's on other material surfaces \cite{bluhm09}, where $\rhoA\sim0.1-10\unit{\mu_B/nm^2}$.  We believe that this is a consequence of the high surface quality of the present samples.  
% suggesting that properties of shallow NV centers can be improved by careful surface preparation.
Given the low density and close proximity of investigated NV centers to the surface, actually only very few surface states significantly couple to the NV spin.  In fact, we have calculated that at a depth of $d\sim3\unit{nm}$ about 80\% of $\Brms^2$ will originate from a single surface impurity.  This means that at shallow depth a single impurity is responsible for spin relaxation.  While this is an exciting prospect in the context of quantum sensing \cite{schaffry11,grinolds13}, it is difficult to confirm and utilize the ``quantum'' character of these surface states due to the short $\tc$.

Noting that $\tc$ and $\rhoA$ do not vary significantly between samples, we have used $\Brms$ to ``gauge'' the approximate depth $d$ (upper bound) of investigated NV centers, according to Eq. (\ref{eq:rho}).
Fig. \ref{fig:data}(e) plots a histogram of inferred depth values.  The histogram suggests that no NV centers lie within 2 nm from the surface.  Although our inferred depth is very approximate, we note that this observation is consistent with the smallest reported size for NV-carrying nanodiamonds of $\oslash\sim 4-5\unit{nm}$ \cite{smith09, tisler09} and recent depth measurements by nanoscale NMR \cite{jelezko13}.

\begin{figure}[t]
\centering
\includegraphics[width=0.48\textwidth]{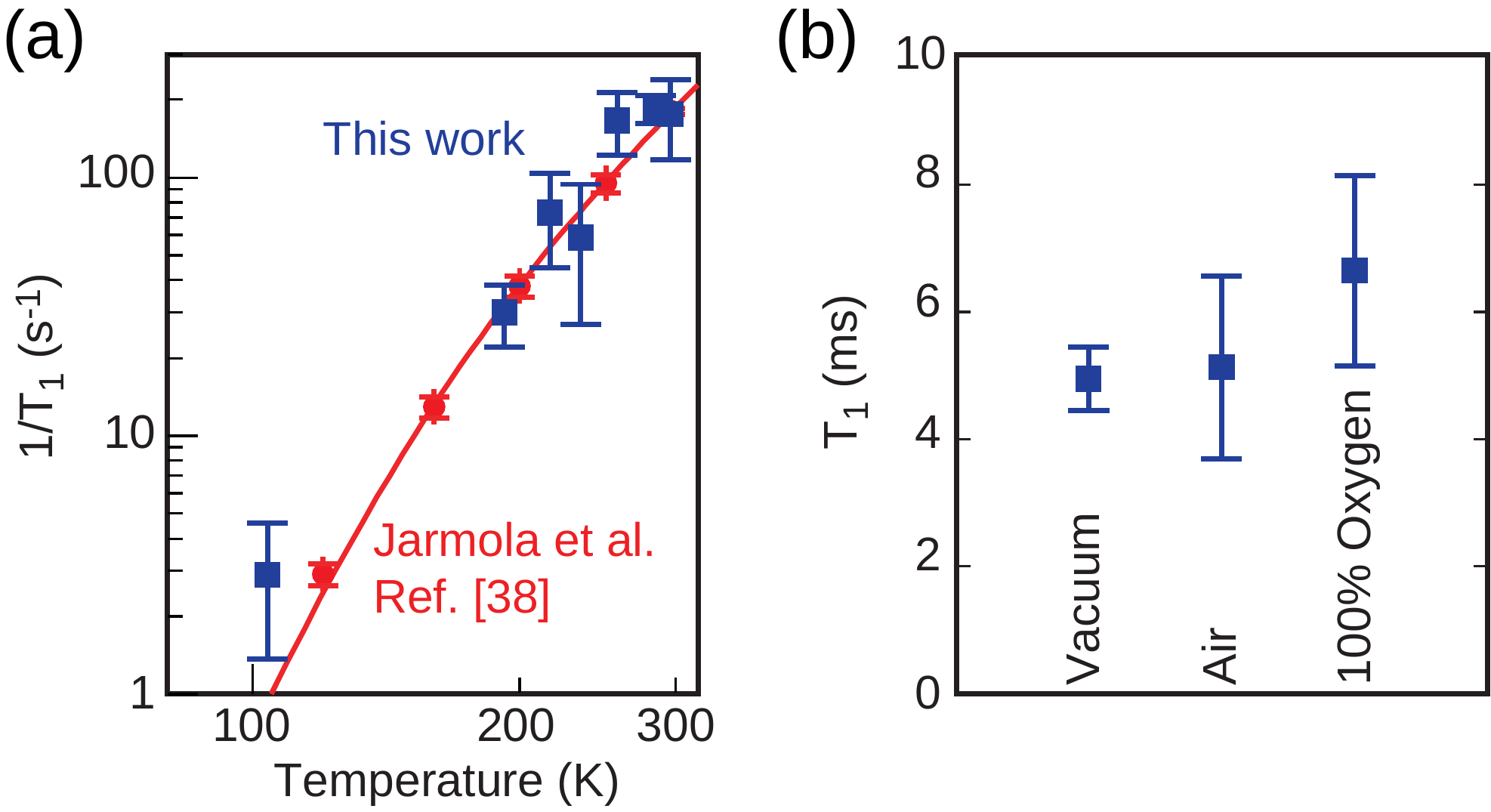}
\caption{\captionstyle
(a) Temperature dependence of $\To$ for a shallow NV center (Sample A, red dots) and for a bulk NV center (from Ref. \cite{jarmola12}, Fig. 2, curve S8).
(b) $\To$ for the same NV center, showing no significant change in $\To$ with exposure to different oxygen pressures.
}
\label{fig:temperature}
\end{figure}

Our data further give insight into the mechanism generating the magnetic fluctuations.  Two main mechanisms have been suggested including spin diffusion and spin-phonon relaxation \cite{desousa07,tetienne13}.
The low density $\rhoA$ of surface states in our samples favors spin-phonon relaxation over spin diffusion.  This hypothesis is supported by the observation that all investigated surfaces show similar correlation times $\tc$ irrespective of $\rhoA$.  To more conclusively establish the mechanism of noise generation we have recorded $\To$ of one NV center as a function of temperature.  Since $\tc\propto\To$ according to Eq. (\ref{eq:SB}) a temperature dependence of $\To$ directly indicates whether fluctuations are thermally activated, as predicted for a spin-phonon (but not a spin diffusion) process \cite{jarmola12}.  
As Fig. \ref{fig:temperature}(a) shows, $\To$ is strongly temperature dependent, indicating that surface fluctuations are indeed thermally activated.  We note that the long $\To$ at low temperature may become benefitial for magnetometry applications that require high frequency resolution \cite{laraoui13}.

We finally discuss a few anecdotal observations.  In an attempt to perform nanoscale NMR measurements with shallow NV centers \cite{mamin13,staudacher13,ohashi13} we have overcoated the diamond surface with a variety of substances and recorded the associated relaxation times.  We did not find any significant changes with any of the substances tested, including stearic acid and optical immersion oils (data not shown).  We found a strong reduction of relaxation times when overcoating the surface by PMMA resist, but this reduction was most likely caused by paramagnetic contamination of the resist.  We have finally exposed the sample to vacuum, ambient air and 100\% oxygen atmospheres, with no noticeable change in $\To$ (see Fig. \ref{fig:temperature}(b)).  We can thus exclude molecular oxygen as the leading cause of surface magnetic noise.
Together, all observations support the general picture where the surface states are intrinsically associated with diamond's top carbon layers \cite{samsonenko79} and not with terminating surface atoms or adsorbate molecules.

In the light of these findings, several illuminating experiments could be conceived to more precisely pinpoint the underlying atomistic mechanism responsible for surface magnetic states.  In particular, different surface orientations of diamond (such as a $(111)$-oriented surface) or atomically-flat substrates \cite{watanabe99} could be explored to elucidate the influence of bonding structure of the top carbon atoms.  %Moreover, the temperature dependence of $\tc$ should reveal whether surface fluctuations are indeed spin-phonon driven.
Altogether, a precise understanding of diamond surface magnetic states will be crucial for further improving the sensitivity and resolution of diamond magnetic sensor heads and sensor arrays.

The work at ETH was supported by the Swiss National Science Foundation through Project Grant $200021\_137520/1$ and through the NCCR QSIT.
The work at Keio has been supported by the Cannon Foundation, the Core-to-Core Program by JSPS, and the Project for Developing Innovation Systems by MEXT.
We thank K. Chang, R. Schirhagl and F. Jelezko for experimental help and fruitful discussions, and J. Meijer and S. Pezzanga for help with preparation of sample B \cite{oforiokai12}.


\begin{thebibliography}{}
\bibitem{depth}
  The thickness of the $\delta$-doped layer was estimated from the calibrated growth rate and is expected to be accurate within 20 \%. See also Refs. \cite{ohashi13,ishikawa12}.
\bibitem{biasfield}
  A small bias field does not significantly alter the relaxation dynamics; the error added to spectral densities is $<2\%$.
\end{thebibliography}
\end{document}